\documentclass[reprint]{revtex4-1}
\pdfoutput=1

\usepackage{graphicx}
\usepackage{amsmath}
\usepackage{amssymb}
\usepackage{chemarr}
\usepackage{textgreek}
\usepackage{mathtools}
\usepackage{commath}
\usepackage{siunitx}
\usepackage[usenames,dvipsnames]{xcolor}

\usepackage[linktocpage=true]{hyperref}
\hypersetup{colorlinks=false}

\usepackage{setspace}
\usepackage{listings}
	
\usepackage[noend]{algpseudocode}

\newcommand{\vect}[1]{\mathbf{#1}}

\usepackage{txfonts}
\usepackage[T1]{fontenc}
\usepackage[scaled]{helvet}
\usepackage[scaled=0.9]{couriers}

\usepackage{titlesec}
\titleformat*{\section}{\large\sffamily\bfseries}
\titleformat*{\subsection}{\sffamily\bfseries}
\titleformat*{\subsubsection}{\small\sffamily\bfseries}

\newcommand*{\titlefont}{\Large\sffamily\bfseries}

\begin{document}

\title{\titlefont Efficient Stochastic Simulation of Chemical Kinetics Networks Using A Weighted Ensemble Of Trajectories}

\author{Rory M. Donovan}

\author{Andrew J. Sedgewick}

\author{James R. Faeder}
\email[]{faeder@pitt.edu}
\homepage[]{http://www.csb.pitt.edu/Faculty/Faeder/}

\author{Daniel M. Zuckerman}
\email[]{ddmmzz@pitt.edu}
\homepage{http://www.csb.pitt.edu/Faculty/zuckerman}

\affiliation{Department of Computational and Systems Biology, University of Pittsburgh}

\date{\today}

\begin{abstract}
We apply the ``weighted ensemble'' (WE) simulation strategy, previously employed in the context of molecular dynamics simulations, to a series of systems-biology models that range in complexity from one-dimensional to a system with 354 species and 3680 reactions.  WE is relatively easy to implement, does not require extensive hand-tuning of parameters, does not depend on the details of the simulation algorithm, and can facilitate the simulation of extremely rare events.

For the coupled stochastic reaction systems we study, WE is able to produce accurate and efficient approximations of the joint probability distribution for all chemical species for all time \(t\).  WE is also able to efficiently extract mean first passage times for the systems, via the construction of a steady-state condition with feedback.  In all cases studied here, WE results agree with independent calculations, but significantly enhance the precision with which  rare or slow processes can be characterized.  Speedups over ``brute-force'' in sampling rare events via the Gillespie direct Stochastic Simulation Algorithm range from \(\sim\)\(10^{12}\) to \(\sim\)\(10^{20}\) for rare states in a distribution, and \(\sim\)\(10^{2}\) to \(\sim\)\(10^{4}\) for finding mean first passage times.  

\end{abstract}

\pacs{87.10.Mn, 87.17.Aa, 87.18.Vf}
\keywords{Stochastic Modeling, Systems Biology, Rare Events, Chemical Networks, Weighted Ensemble}

\maketitle

\section{Introduction}

Stochastic behavior is an essential facet of biological processes such as gene expression, protein expression, and epigenetic processes~\cite{Esteller2008,McAdams1997,Arkin1998,Blake2003,Raser2004,Weinberger2005,Acar2008,Cai2006,Elowitz2002,Raj2008,Maheshri2007,Kaufmann2007,Shahrezaei2008,Raj2009}.  Stochastic chemical kinetics simulations are often used to study systems biology models of such processes~\cite{Gillespie2007,Wilkinson2011,Blinov2004}.  One of the more common stochastic approaches, and the one employed in the present study, is the stochastic simulation algorithm (SSA), also known as the Gillespie algorithm~\cite{Gillespie1976,Gillespie1977,Gillespie2007}.

As stochastic systems biology models approach the true complexity of the systems being modeled, it quickly becomes intractable to investigate rare behaviors using na\"ive (``brute-force'') simulation approaches.  By their very nature, rare events occur infrequently; confoundingly, rare events are often those of most interest. For example, the switching of a bistable system from one state to another may happen so infrequently that running a stochastic simulation long enough to see transitions is (extremely) computationally prohibitive~\cite{Allen2005}.  This impediment only grows as model complexity increases, and as such it poses a serious hurdle for systems models as they grow more intricate.

Several approaches to speeding up the simulation of rare events in stochastic chemical kinetic systems exist.  A variety of ``leaping'' methods can, by taking advantage of approximate time-scale separation, accelerate the SSA itself~\cite{Zhou2008,Mjolsness2009,Jenkins2011,Chatterjee2005,Bayati2009,Gillespie2003,Lu2012,Gibson2000}.  Kuwahara and Mura's weighted stochastic simulation (wSSA) method~\cite{Kuwahara2008} was refined by Gillespie and Petzold et al.~\cite{Gillespie2009,Roh2010,Daigle2011,Roh2011}, and is based on importance sampling.  The forward flux sampling method of ten Wolde et al.~\cite{Allen2005,Allen2006a,Allen2006,Allen2009} uses a series of interfaces in state-space to reduce computational effort, as does the non-equilibrium umbrella sampling approach~\cite{Dickson2009,Warmflash2007}.

Rare event sampling is also an active topic in the field of molecular dynamics simulations, and many approaches have been proposed.  Of the approaches that do not irreversibly modify the free energy landscape of the system, some notable methods include dynamic importance sampling~\cite{Zuckerman1999}, milestoning~\cite{Faradjian2004}, transition path sampling~\cite{Dellago1998}, transition interface sampling~\cite{VanErp2003}, forward flux
sampling~\cite{Allen2009}, non-equilibrium umbrella sampling~\cite{Warmflash2007}, and weighted ensemble sampling~\cite{Huber1996,Zhang2007,Zhang2010,Bhatt2010,Zwier2011,Adelman2013,Lettieri2012,Zwier2013}.  For a summary of these methods, see~\cite{Zwier2010}.  Many of the ideas behind these techniques are not exclusive to molecular dynamics simulations, and can be adapted to studying stochastic chemical kinetic models.  For example, dynamic importance sampling seems to be closely related to wSSA.

Because of its relative simplicity and potential simplicity in sampling rare events, we apply one of these methods, the weighted ensemble algorithm (WE) to well-established model systems of stochastic kinetic chemical reactions.  These models range in complexity from one species and two reactions, to 354 species and 3680 reactions.  For the systems studied, WE proves many orders of magnitude faster than SSA simulation alone, offers linear parallel scaling, returns full distributions of desired species at arbitrary times, and can yield mean first passage times (MFPTs) via the setup of a feedback steady-state.

\section{Methodology}

The methods employed are described immediately below, while the models are specified in Sec.\ \ref{sec:modelsandresults}.

\subsection{Stochastic Chemical Kinetics \& BioNetGen}

Stochastic chemical kinetics occupies a middle-ground in the realm of chemical simulation, between very explicit, and costly, molecular dynamics (MD) simulations and the deterministic formalism of reaction rate equations (RRE).  Stochastic chemical kinetics attempts to account for the randomness inherent in chemical reactions, without trying to explicitly model the spatial structure of the reacting species.  It is many orders of magnitude faster than MD simulations, but much slower than the RRE approach.  It is an ideal method to use for modeling the effects of low concentrations (or copy numbers) of chemical reactants, while ignoring the effects of specific spatial distribution.  

Stochastic chemical kinetics models can be solved exactly for sufficiently simple systems using the Chemical Master Equation (CME), and approximately (for all systems) using Gillespie's direct stochastic simulation algorithm (SSA)~\cite{Gillespie1976,Gillespie1977,Gillespie2007}.  The SSA samples the CME exact solution by modeling stochastic chemical kinetics in a straightforward manner, and yields trajectories of species concentrations that converge to the RRE method in the limit of large amounts of reactants.  In brief, the SSA iteratively and stochastically determines which reaction fires when reactions occur, by sampling from the exponential distribution of waiting times between reactions.  For a detailed explanation of the SSA, see \cite{Gillespie2007}.

We employ the rule-based modeling and simulation package BioNetGen~\cite{Faeder2009} to simulate both our toy and complex models.  Rule-based modeling languages allow the specification of biochemical networks based on molecular interactions.  Rules that describe those interactions can be used to generate a reaction network that can be simulated either as RREs or using the SSA, or the rules can be used directly to drive stochastic chemical kinetics simulations.  BioNetGen has been applied to a variety of systems, such as the aggregation of membrane proteins by cytosolic cross-linkers in the LAT-Grb2-SOS1 system~\cite{Nag2009}, the single-cell quantification of IL-2 response by effector and regulatory T cells~\cite{Feinerman2010}, the analysis and verification of the HMGB1 signaling pathway~\cite{Gong2010}, the role of scaffold number in yeast signaling systems~\cite{Thomson2011}, and the analysis of the roles of Lyn and Fyn in early events in B cell antigen receptor signaling~\cite{Barua2012}.  We employ BioNetGen's implementation of the direct SSA to propagate the dynamics in our systems.  

\subsection{Weighted Ensemble (WE)}

WE is a general-purpose protocol used in molecular dynamics simulations~\cite{Zhang2007,Zhang2010,Bhatt2010,Adelman2013,Lettieri2012,Zwier2013}
that we adapt here to the efficient sampling of dynamics generated by chemical kinetic models. In brief, WE employs a strategy of ``statistical natural selection'' using quasi-independent parallel simulations which are coupled by the intermittent exchange of information. The intermittency leads directly to linear parallel scaling.  Importantly, the simulations are coupled via configuration space (essentially the ``phase space'' of the system in physics language or the ``state-space'' in cell and population modeling).  This type of coupling permits both efficiency and a large degree of scale independence. The efficiency results from distributing trajectories to typically under-sampled parts of the space, while scale independence is afforded because every type of system has a configuration or state-space.

\begin{figure*}[htbp]
   \centering
   \includegraphics[width=17cm]{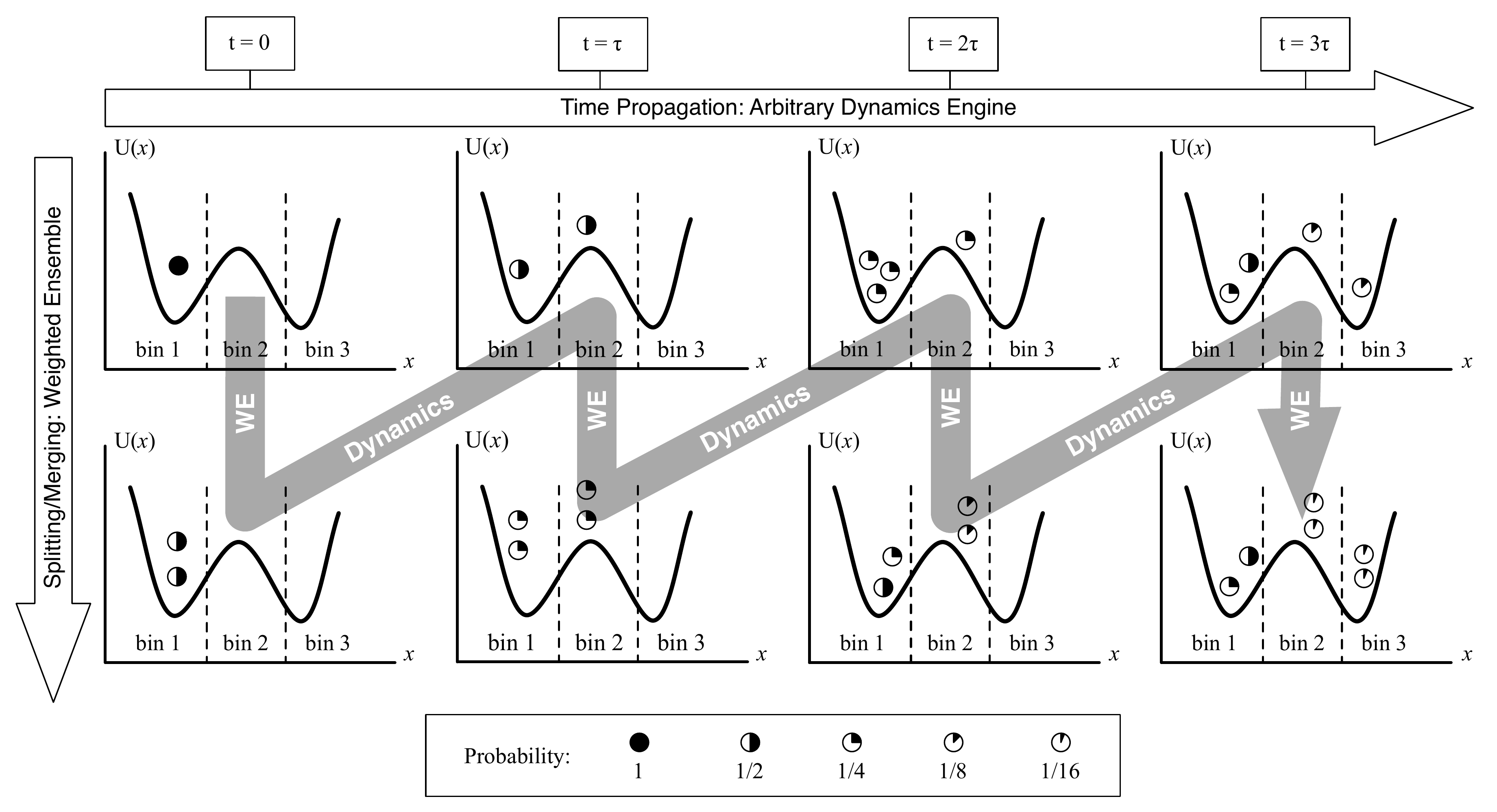} 
   \caption[Weighted Ensemble Schematic Description]{Weighted ensemble (WE) simulation depicted for a configuration/state-space divided into bins. Multiple trajectories are run using any dynamics software (here we use the SSA in BioNetGen) and checked every \(\tau\) for bin location. Trajectories are assigned weights (symbols -- see legend) that sum to one and are split and combined according to statistical rules that preserve unbiased kinetic behavior.}
   \label{fig:WEcartoon}
\end{figure*}

WE's strategy of statistical natural selection or statistical ratcheting is schematized in Fig.~\ref{fig:WEcartoon}. First, the space is divided/classified into non-overlapping ``bins'' which are typically static, although dynamic and adaptive tessellations are possible~\cite{Zhang2010}. A target number of trajectories,~\(M_\text{targ}\), is set for each bin. Multiple trajectories are initiated and each is assigned a weight so that the sum of weights is one. Trajectories are then simulated independently according to the desired dynamics (e.g., molecular dynamics or SSA) and checked intermittently (every \(\tau\) units of time) for their location. If a trajectory of weight \(w\) is found to occupy a previously unoccupied bin, that trajectory is replicated to obtain the target number of copies, \(M_\text{targ}\), for the bin. Daughter trajectories' weights are set to \(w/M_\text{targ}\), to sum to the weight of the parent trajectory. If a bin is occupied by more than the target number, trajectories must be pruned in a statistical fashion maintaining the sum of weights. Specifically, the two lowest weight trajectories are ``merged'' by randomly selecting one of them to survive, with probability proportional to their weights, and the surviving trajectory absorbs the weight of the pruned one.  This process is repeated as needed, and maintains an exact statistical representation of the evolving distribution of trajectories~\cite{Zhang2010}.

Setting up a WE simulation requires selection of state-space binning, trajectory multiplicity, and timing parameters.  In our simulations, we chose to divide the state-space of an \(N\)-dimensional system into one- or two-dimensional regular grids of non-overlapping bins.  It is possible to use non-Cartesian bins, and to adaptively change the bins during simulation~\cite{Zhang2010,Adelman2013}, but for simplicity we did not pursue any such optimization.  Specific parameter choices for each model are given in Sec.\ \ref{sec:modelsandresults}.

The weighted ensemble algorithm can be outlined fairly concisely.  Let \(M_\text{targ}\) be the target number of segments in each bin, \(N_\text{bins}\) the number of bins, whose geometry are defined by the grid \(G_\text{grid}\), \(\tau\) the time-step of an iteration of WE, and \(N_\text{iters}\) the total number of iterations of WE.  The WE procedure also requires an initial state of the system, \(\vect{x}_0\), which in our case is a list of the concentrations of all the chemical species in the system.
\begin{algorithmic}
\Procedure{WE}{$N_\text{iters},\tau,G_\text{grid},M_\text{targ},\vect{x}_0$}
	\For{$i = 1 \ldots N_\text{iters}$}
		\For{each populated bin in $G_\text{grid}$}		
			\State{propagate dynamics for all trajectories	}
		\EndFor
		\For{each bin in $G_\text{grid}$}
			\If{bin population $= 0$ or $M_\text{targ}$}
				\State do nothing
			\ElsIf{bin population $< M_\text{targ}$}
				\State replicate trajectories until bin pop.\ $=M_\text{targ}$
				\State maintain sum of weights in each bin
			\ElsIf{bin population $> M_\text{targ}$}
				\State merge trajectories until bin pop.\ $=M_\text{targ}$
				\State maintain sum of weights in each bin
			\EndIf
		\EndFor
		\State{save coordinates and weights of each trajectory}
	\EndFor
	\State \textbf{return} trajectory coordinates \& weights for each iter.
\EndProcedure
\end{algorithmic}
The replicating and merging of trajectories in the above algorithm are done randomly, according to the weight of each trajectory segment in a given bin, which has been shown not to bias the dynamics of the ensemble~\cite{Huber1996,Bhatt2010}.

When WE is used to manage an ensemble of trajectories, there are two time-scales of immediate concern: the period at which trajectory coordinates are saved, and the period \(\tau\) at which  ensemble operations are performed.  These two time-scales can be different, but for simplicity we set them to be the same, and select \(\tau\) such that it is greater than the average event firing rate for the SSA.  When we refer to the time-step, or iteration of a process, we are referring to the \(\tau\) of Fig.~\ref{fig:WEcartoon}.

WE can be employed in a variety of modes to address different questions. Originally developed to monitor the time evolution of arbitrary initial probability distributions~\cite{Huber1996}, i.e.\ non-stationary non-equilibrium systems, WE was generalized to efficiently simulate both equilibrium and non-equilibrium steady-states~\cite{Bhatt2010}. In steady-state mode, mean first passage times (MFPTs) can be estimated rapidly based on simulations much shorter than the MFPT using a simple rigorous relation between the flux and MFPT~\cite{Bhatt2010}.  Steady-states can be attained rapidly, avoiding long relaxation times, by using the inter-bin rates computed during a simulation to estimate bin probabilities appropriate to the desired steady-state; trajectories are then reweighted to conform to the steady-state bin probabilities~\cite{Bhatt2010}.  Both of these methods are described in more detail below.

\subsubsection{Basic WE: Probability Distribution Evolving in Time}

Perhaps the simplest use of a weighted ensemble of trajectories is to better sample rare states as a system evolves in time, specifically the states corresponding to extreme values of the binning coordinate.  The SSA itself samples the exact distribution, but its sampling is concentrated about the mode(s) of the distribution.  The SSA naturally -- and correctly -- samples rare states infrequently.  By using WE to split up the state-space, however, one can resample the distribution at every time step \(\tau\), selecting those trajectories that advance along a progress coordinate for more detailed study, but doing so without applying any forces or biasing the trajectories or the distribution.  Essentially, WE appropriates much of the effort that brute-force SSA devotes to sampling the central component of the distribution, repurposing it to obtain better estimates of the tails.

This basic use of WE requires none of the ``tricks'' we apply in later sections, such as using reweighting techniques to accelerate obtaining a steady-state.  We apply basic WE to some of our systems -- particularly, but not exclusively, to those that are not bistable.

\subsubsection{Steady-State}

The mean first passage time (MFPT) from state \(A\) to state \(B\) is a key observable.  It is equal to the inverse of the flux (of probability density) from state \(A\) to state \(B\) in steady-state \cite{Hill2004},
\begin{equation}
\text{MFPT}_{A \to B} = \frac{1}{\text{Flux}_{ss}(A \to B)} \: .
\label{eqn:mfpt}
\end{equation}
This relation provides the weighted ensemble approach the ability to calculate MFPTs in a straightforward manner.  During a WE run, when any trajectories (and their associated weights) reach a designated target area of state-space (or ``state \(B\)''), they are removed and placed back in the initial state (``state \(A\)'').  Eventually, such a process will result in a steady-state flow of probability from state \(A\) to state \(B\) that does not change in time (other than with stochastic noise).

\paragraph*{Reweighting.} \label{par:reweighting} The waiting time to obtain a steady-state constrains the efficiency of obtaining a MFPT by measuring fluxes via equation \ref{eqn:mfpt}.  This waiting time can vary from the relatively short time scale of intra-state equilibration for simple systems, to much longer time-scales, on the order of the MFPT itself for more complicated systems.  To reduce this waiting time, we use the steady-state reweighting procedure of Bhatt et al.~\cite{Bhatt2010}.  This method measures the fluxes between bins to obtain a rate-matrix for transitions between bins, and uses a Markov formulation to infer a steady-state distribution from the (noisy) data available.

For instance, let \(\{w_i\}\) be the set of bin weights (i.e\ the sum of the weights of the trajectories in each bin), and let \(\{w_i^{ss}\}\) be the set of steady-state values of the bin weights.  If \(f_{ij}\) is the flux of weight into bin \(i\) from bin \(j\), then in steady-state, since the flux out of a bin is equal to the flux into it,
\begin{equation}
\label{eq:ss}
\frac{\dif w_i^{ss}}{\dif t} = \sum_j\left( f_{ij}^{ss} -f_{ji}^{ss} \right) = \sum_j\left( k_{ij} w_j^{ss} - k_{ji} w_i^{ss} \right) = 0 \;.
\end{equation}
Since the flux of weight into bin \(i\) from bin \(j\) is the product of a (constant) rate and the (current) weight in a bin, i.e.\ \(f_{ij}~=~k_{ij}~w_j\) (true for both steady state and not), we can use Eq.\ \ref{eq:ss} to find the inter-bin rates.  By measuring the inter-bin fluxes and the bin weights, we can approximately infer the transition rates, and then find a set of weights that satisfy Eq.\ \ref{eq:ss}.  Once the set of bin weights is found, the weights of the individual trajectories in the bins are rescaled commensurately.

The steady-state distribution of weights thus inferred is not necessarily the true steady-state of the system, but it tends to be closer to it, and an iterative application of this procedure can converge to the true distribution fairly rapidly.  In practice it has been shown to accelerate the system's evolution to a true steady-state by orders of magnitude in some cases~\cite{Bhatt2010}.

\subsection{Estimation of Computational Efficiency}
\label{subsec:efficiency}

Since it is important to assess new approaches quantitatively, we compare the speedup in computing time from weighted ensemble to a brute-force simulation, (i.e.\ SSA).  For a given observable (e.g., the fraction of probability in a specified tail of the distribution) and a desired precision, we estimate the efficiency using the ratio:
\begin{equation}
E \coloneqq \frac{\text{dynamics time in brute-force SSA}}{\text{dynamics time in WE-SSA}}
\end{equation}
Since both WE and brute-force use the same dynamics engine/software, we can estimate the speedup of WE over brute-force by just keeping track of how much total ``dynamics time'' was simulated in each.  We employ this measure when estimating the advantage of using WE to investigate the tails of probability distributions, as well as for finding MFPTs in bistable systems.

Another measure of efficiency we employ for MFPT estimation gauges how fast WE attains a result that is within 50\% of the true result (determined from exact or extensive brute-force calculation):  
\begin{equation}
E_{50\%} \coloneqq \frac{\text{dynamics time in brute-force SSA to get} \pm 50\% \text{ exact}}{\text{dynamics time in WE-SSA to get}\pm 50\% \text{ exact}} \: .
\end{equation}
This is an assessment of how well WE can extract rough estimates of long time-scale behavior from simulations that are much shorter than those timescales.  

Brute-force SSA simulations can be run for long times without seeing a transition from one macro-state to another.  To take account of the brute-force simulations where no transitions occurred we use a maximum likelihood estimator for the transition time, based on an exponential distribution of waiting times, which is a valid approximation for the one-dimensional and two-state systems studied below:

\begin{equation}
\label{eq:mld}
\begin{aligned}
\mu_{MLE} &= \left( 1 - \frac{n}{N} \right) T + \frac{1}{n}\sum_{i=1}^n t_i \\
\sigma_\mu &= \frac{\mu_{MLE}}{\sqrt{n}}
\end{aligned}
\end{equation}
where \(T\) is the length of the brute-force simulations, \(N\) is the number of these simulations performed, \(n\) is the number of these simulations in which a transition from one state to another is observed, and \(t_i\) are the times at which the transition is observed.

\subsection{Limitations of Our Implementation}

We used two different implementations of the weighted ensemble framework: WESTPA, written in Python, is the most feature-rich and stable~\cite{Zwier2013}, which will be available at \url{http://chong.chem.pitt.edu/WESTPA}.  Another, written by Bin Zhang~\cite{Zhang2007} and modified by us, is written in C, and is faster though less robust, and is available at \url{http://donovanr.github.com/WE_git_code}.

Weighted ensemble (WE), as a scripting-level approach, inherently adds some unavoidable overhead to the runtime of the dynamics.  This overhead, in theory, is quite minimal: stopping, starting, merging, and splitting trajectories are not computationally costly operations.  A key issue in practical implementations, though, is how long the algorithm actually takes to run, i.e\ the wall-clock running-time for dynamics (here, the SSA).

In practice, overhead can be significant for very simple systems, for the sole reason that reading and writing to disk takes so much time compared to how long it takes to run the dynamics of small models.  In our implementation, data is passed from the dynamics engine to WE by reading and writing files to disk.  This handicap is an artifact of our interface, which could, with minimal work, be modified to something more efficient.  As a proof-of-principle, the version of WE written in C was modified, for the Schl\"ogl reactions and the futile cycle, to contain hard-coded versions of the Gillespie direct algorithm for those systems, so as to obviate the I/O between WE and BNG.  With these modifications, it was difficult to ascertain any significant overhead costs at all, and our runs completed in a matter of seconds.  We also note that as model complexity increases and more time is devoted to dynamics, the overhead problem becomes negligible. Practical applications of WE will, by nature, target models where dynamics are expensive, rather than toy models, where they are cheap.

\section{Models \& Results}

\label{sec:modelsandresults}

We study four different models, ranging in complexity from two chemical reactions governing one chemical species, to 3680 reactions governing 354 species.  The models we employ are coupled stochastic chemical reactions, which we implement and simulate in BioNetGen using the SSA~\cite{Gillespie1976,Gillespie1977,Gillespie2007}.  As depicted in Fig.~\ref{fig:WEcartoon}, these simulations are, in turn, managed by a weighted ensemble procedure.

\subsection{Enzymatic Futile Cycle}

\subsubsection{Model}

The enzymatic futile cycle is a simple and robust model that can, in certain parameter regimes, exhibit qualitatively different behavior due to stochastic noise~\cite{Samoilov2005,Warmflash2008}. This signaling motif can be seen in biological systems including GTPase cycles, MAPK cascades, and glucose mobilization~\cite{Samoilov2005,Kholodenko2006,Wang2008}.

The enzymatic futile cycle studied here is modeled by:
\begin{equation}
\begin{aligned} \label{eq:futile}
E_1 + S_1 \xrightleftharpoons[k_f]{k_f} B_1 \xrightarrow{k_s} E_1 + S_2 \\
E_2 + S_2 \xrightleftharpoons[k_f]{k_f} B_2 \xrightarrow{k_s} E_2 + S_1
\end{aligned}
\end{equation}
where \(k_f=1.0\) and \(k_s=0.1\).  Here \(S_1\) can bind to its enzyme \(E_1\), and in the bound form, \(B_1\), (i.e.\ \( B_1=E_1 \cdot S_1 \)), it can be converted to \(S_2\), and then dissociate (and similarly for \(S_2 \xrightarrow{} S_1\)).  The total amount of substrate, \(S_1 + S_2\), is conserved, as are the amounts of the different enzymes \(E_1\) and \(E_2\), of which is supplied only one of each kind.  Following Kuwahara and Mura~\cite{Kuwahara2008}, in the specific system we look at, we set \(S_1 + S_2 = 100\) and \(E_1 + B_1 = E_2 + B_2 = 1\).

Thus constrained, the above system of reactions can be solved by a 404-state chemical master equation (CME), to obtain an exact probability density for all times when initialized from an arbitrary starting point.  We start the system at \(S_1 = S_2 = 50\) and \(E_1 = E_2 = 1\), and are interested in the probability distribution of \(S_1\) after 100 seconds, that is, \(P(S_1=x,t=100)\).

\subsubsection{WE Parameters}

The WE data was generated using 101 bins of unit width on the coordinate \(S_1\).  We employed 100 trajectory segments per bin that were run for 100 iterations of a \(\tau = 1 \: \si{\second}\) time-step, with no reweighting events.  The brute-force data is from 10,200 100-second runs, which is an equivalent amount of dynamics to compute as the single WE run, if all the bins were full all the time.  However, since the bins take some time to fill up, the WE run employed only 840,000 one-second segments, which makes the comparison to brute-force SSA more than fair.

\subsubsection{Results}
\label{sec:futileresults}

\begin{figure}[htbp]
   \centering
   \includegraphics[width=8.5cm]{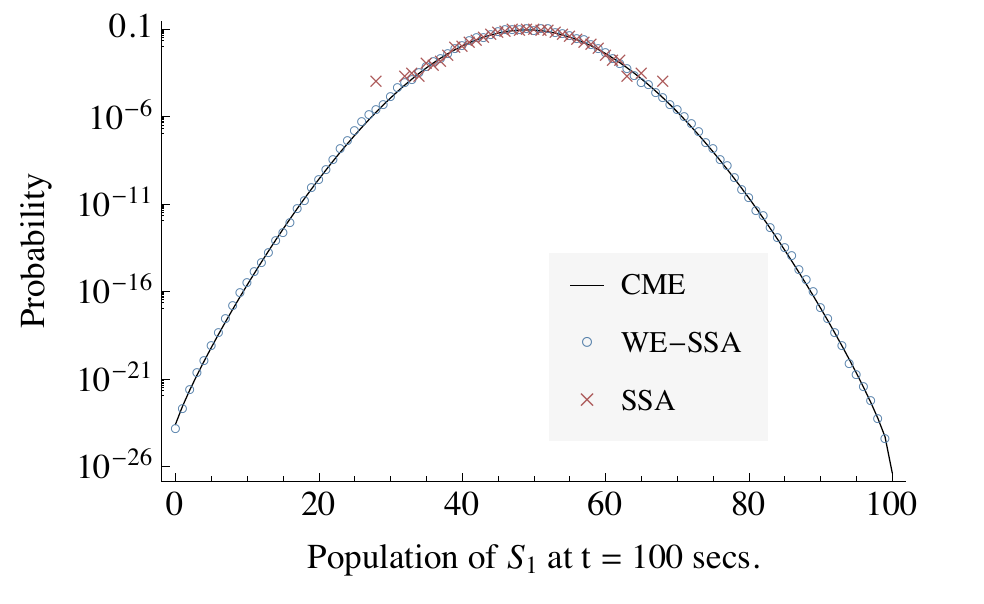} 
   \caption[Enzymatic Futile Cycle PDF]{The probability distribution of \(S_1\) in the Enzymatic Futile Cycle System, after \(t=100\) seconds, when initialized from a delta function at \(S_1=50\), \(E_1=E_2=1\) at \(t=0\).  The exact solution, procured via the chemical master equation (CME), is compared to data obtained using the SSA in a weighted ensemble run (WE-SSA), and to ordinary SSA, when each are given equal computation time.  WE data is from a single run.  Error bars are not plotted; for a discussion of uncertainties, see Sec.~\ref{sec:futileresults}.}
   \label{fig:futilePDF}
\end{figure}

Fig.~\ref{fig:futilePDF} shows that the brute-force SSA is unable to sample values of \(S_1\) much outside the range \(30 < S_1 < 70\), whereas the WE method is able to accurately sample the entire distribution.  Waiting for the brute-force approach to sample the tails would take \(\sim 1/P(\text{tail}) \sim 1/10^{-23} \sim 10^{23}\) brute-force runs.  With a conservative estimate of \(\sim\)\( 10^{4}\) runs per second, it would take \(\sim\)\( 10^{19}\) seconds, or many times the age of the universe, for brute-force SSA to sample the tails at all.  WE takes 2--3 seconds to sample them  (note the comparison to exact distribution provided by the CME), for an approximate efficiency increase \(E \sim 10^{18}\).

For the sake of clarity, error-bars were omitted from Fig.~\ref{fig:futilePDF}.   Over most of the data range, the error is too small to see on the plot. In the tails (of both SSA and WE-SSA) the error is not computable from a single run, since there are plot points comprised of only a single trajectory.  The error in the estimate of the distribution can be inferred visually from the data's departure form the CME exact solution.  For SSA, however, generating uncertainties far all values is essentially impossible.  When computing quantitative observables reported below, we employ multiple independent runs to procure standard errors in our estimate.

From the distribution, we are able to read off useful statistics.  Instead of computing MFPTs, Kuwahara and Mura~\cite{Kuwahara2008}, and Petzold and Gillespie et al.~\cite{Gillespie2009} defined a related quantity, the probability of a system to pass from one state to another in a certain time: \( P(x_i \rightarrow x_f | \Delta t) \)~\cite{Kuwahara2008,Gillespie2009,Roh2010,Daigle2011,Roh2011}.  For instance, one might desire to know the probability of the futile cycle to have a value of \(S_1 > 90\) at \(\Delta t=100\).  Since WE gives an accurate estimate of \(P(x,t)\), all that is required to find such statistics is to sum up the area under the state of interest.  For the futile cycle, we find a value of \(2.47\times10^{-18} \pm 3.4\times10^{-19}\) at one standard error for \(P(S_1 > 90, t=100 \: \si{\second})\),  as computed from ten replicates of the single WE run plotted in Fig.~\ref{fig:futilePDF}.  The CME equation gives an exact value of \(2.72\times10^{-18}\).  Sampling this tail of the distribution at all by brute-force would take \(\sim 1/P(\text{tail}) \sim 1/(2.72\times10^{-18}) \sim 4 \times 10^{17}\) brute-force runs.  Using ten replicate runs, WE is able to sample it using 8,400,000 WE segments, which is equivalent to 83,170 brute-force trajectories, resulting in an increase in sampling efficiency by a factor of \(E \sim 4 \times 10^{17} / 83,170 \sim \times 10^{12}\) for this observable.

\subsection{Schl\"{o}gl Reactions}

\subsubsection{Model}

The Schl\"{o}gl reactions are a classic toy-model for benchmarking stochastic simulations of bistable systems~\cite{Vellela2009}.  They are two coupled reactions with one dynamic species, \(X\): 
\begin{equation}
\begin{aligned} \label{eq:schlogl}
A + 2X &\xrightleftharpoons[k_2]{k_1} 3X \\
B &\xrightleftharpoons[k_4]{k_3} X
\end{aligned}
\end{equation}
where \(k_1=3 \times 10^{-7}\), \(k_2=10^{-4}\), \(k_3=10^{-3}\), \(k_4=3.5\), \(A=10^5\), and \(B=2 \times 10^5\).  The species \(A\) and \(B\) are assumed to be in abundance, and are held constant.  Both the mean first passage times and the time-evolution of arbitrary probability distributions can be computed exactly~\cite{Gillespie1992}.

\subsubsection{WE Parameters}

The WE data in Fig.~\ref{fig:schloglPDF} was generated using 802 bins of unit width, 100 trajectory segments per bin, a time-step \(\tau = 0.05 \: \si{\second}\), and run for 101 iterations of that time-step, with no reweighting events.  The brute-force data is from 80,200 5-second runs, which is an equivalent amount of dynamics as a single WE run, if all bins are always full.  Were that the case, the WE run would compute dynamics for 8,100,200 trajectory segments; in our case the WE simulation ran 7,047,300 trajectory segments, which makes the comparison to brute-force more than fair.

The WE data in Fig.~\ref{fig:schloglFLUX} was generated using 80 bins of width 10, with 32 trajectory segments per bin, a time-step \(\tau = 0.1 \: \si{\second}\), run for 500 iterations of that time-step.  Reweighting events (see Sec.\ \ref{par:reweighting}) were applied every 100, 5, and 2 iterations for the data labeled ``RW-100'', ``RW-5'', and ``RW-2'', respectively.

\subsubsection{Results}
\label{sec:schloglresults}

\begin{figure}[htbp]
   \centering
   \includegraphics[width=8.5cm]{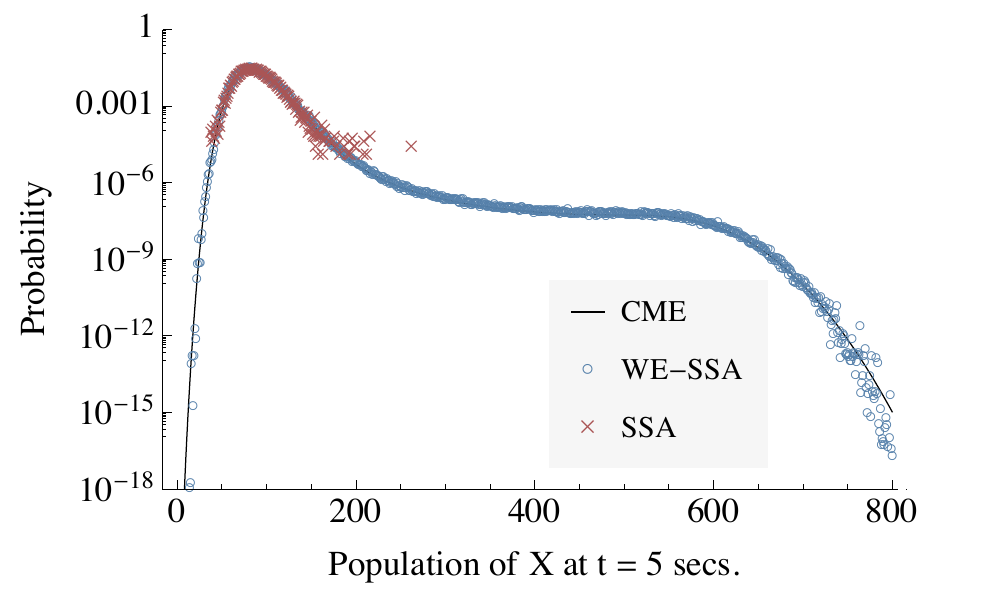}
   \caption[Schl\"ogl Reactions PDF]{The probability distribution of \(X\) in the Schl\"{o}gl system, at \(t=5\) seconds, when initialized from a delta function at \(X=82\).  The exact solution from the chemical master equation is compared to data obtained using the SSA in a weighted ensemble run (WE-SSA), and to ordinary SSA.  WE data is from a single run.  For a discussion of uncertainties, see Sec.~\ref{sec:schloglresults}.}
   \label{fig:schloglPDF}
\end{figure}

Fig.~\ref{fig:schloglPDF} shows how the results of both a brute-force (BF) approach, and the WE approach compare to the exact solution~\cite{Gillespie1992}, when each employs the same amount of dynamics time.  We start the Schl\"{o}gl system with \(X=82\), i.e.\ the PDF is initially a delta function at \(X=82\). To investigate rare transitions, we study the PDF at time \(t=5 \: \si{\second}\).  WE is able to accurately sample almost the entire distributions, even over the potential barrier near \(X=250\), while the BF approach is limited to sampling only high probability states.  The Schl\"ogl system is bistable, with states centered at \(X=82\) and \(X=563\), and a potential barrier between them, peaked at \(X=256\). The brute-force approach is unable to accurately sample values outside of the initial state, and cannot detect bistability in the model.

For the sake of clarity, error-bars were omitted from Fig.~\ref{fig:schloglPDF}.   Over most of the data range, the error is too small to see on the plot. In the tails (of both SSA and WE-SSA) the error is not computable from a single run, since there are plot points comprised of only a single trajectory.  Multiple runs are consistent with the data shown.  The error in the estimate of the distribution can be inferred visually from the data's departure form the CME exact solution.  When computing quantitative observables below, we employ multiple independent runs to procure standard errors in our estimate.

WE yields the full, unbiased probability distribution, but we again examine an observable investigated by Petzold and Gillespie et al.~\cite{Kuwahara2008,Gillespie2009,Roh2010,Daigle2011,Roh2011}.  As such, the conversion to the rare event statistics of Gillespie et al.\ is a simple summation.  From ten replicates of the Schl\"{o}gl run plotted in Fig.~\ref{fig:schloglPDF}, the probability that \(X~\ge~700\) at \(t~=~5\) seconds, i.e.\ \(P(X~\ge~700,~t~=~5 \: \si{\second})\), is \(1.143~\times~10^{-9} \pm 4.7\times~10^{-11}\) at 1-\(\sigma\).  The CME exact value is \(1.148\times10^{-9}\).  Since it would take at least $1 / (1.15~\times~10^{−9}) \sim 10^9$ brute-force trajectories to sample the probability that \(X~\ge~700\) at \(t=5\), we can estimate an improvement in efficiency of using WE over brute-force of \(10^9/802,000 \sim 10^3\).

\begin{figure}[htbp]
   \centering
   \includegraphics[width=8.5cm]{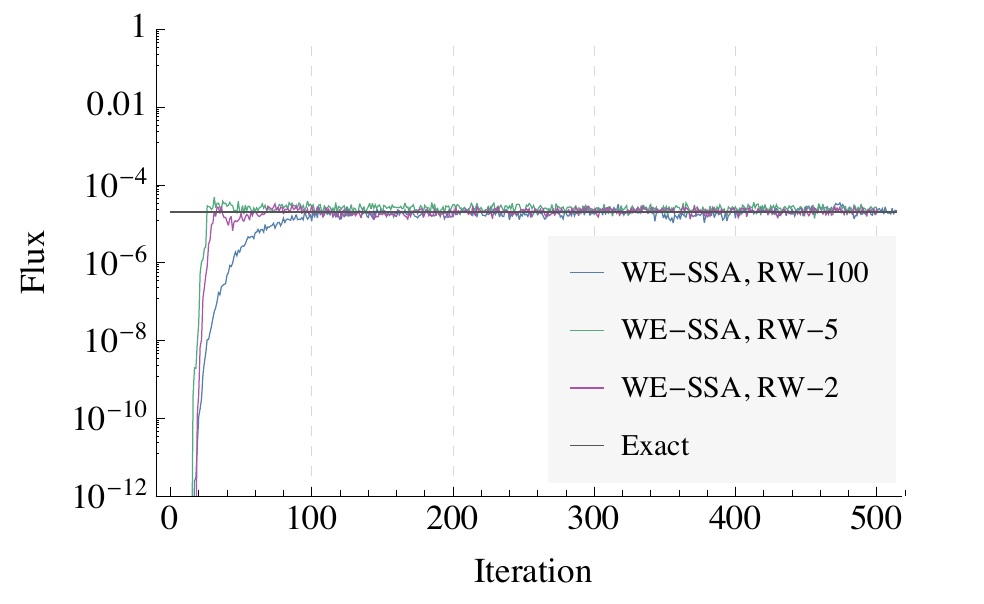} 
   \caption[Schl\"ogl Reactions Flux]{The flux of probability into the target state (\(X \ge 563\)) for the Schl\"{o}gl system.  The exact value is compared to WE results, for reweighting periods of every 100, 5, and 2 iterations.  The inverse of the flux gives the mean first passage time by Eq.\ \protect{\ref{eqn:mfpt}}.}
   \label{fig:schloglFLUX}
\end{figure}

We also estimate the mean first passage time (MFPT) of the Schl\"ogl system, which can be computed exactly~\cite{Gillespie1992}.  Weighted ensemble can estimate the MFPT using Eq.\ \ref{eqn:mfpt} when the system is put into a steady-state.  For the run that was reweighted every 100 iterations, Fig.~\ref{fig:schloglFLUX} shows the WE estimates of the flux from the initial state (\(X=82\)) to the final state (\(X\ge563\)) converge to the exact value in about 100 iterations of weighted ensemble splittings and mergings, which is when the system relaxes from its delta-function initialization to a steady-state.  The attainment of steady-state is accelerated by more frequent reweighting (see Sec.~\ref{par:reweighting} on reweighting), as is shown in Fig.~\ref{fig:schloglFLUX} in the runs that are reweighted every 2 and 5 iterations.  These more frequently reweighted runs yield fluxes close to the exact value within about 30 iterations.

To quantify WE's improvement over brute-force in the estimate of the MFPT, we use the measure \(E_{50\%}\) defined in Sec.\ \ref{subsec:efficiency}.  A brute-force estimate of the MFPT would require, optimistically, computing an amount of dynamics on the order of the MFPT itself (approximately \(5 \times 10^4\)) seconds.  Since transitions in this system follow an exponential distribution, the standard deviation of the first passage times is equal to the mean of them.  WE's estimate of the MFPT is within \(50\%\) of the exact value after about 30 iterations of WE simulation, at which point about 1100 trajectory segments have been propagated, which is equivalent to propagating about 110 seconds of brute-force dynamics.  Thus we find \(E_{50\%} \sim 5 \times 10^4 / 110 \approx 500\).  As can be seen in Fig.~\ref{fig:schloglFLUX}, this value is about a 3--5 fold increase over the WE results when reweighting very infrequently (every 100 iterations).

\subsection{Epigenetic Switch}

\subsubsection{Model}

This model consists of two genes that repress each other's expression. Once expressed, each protein can bind particular DNA sites upstream of the gene which codes for the other protein, thereby repressing its transcription~\cite{Roma2005}. If we denote the \(i\)th protein concentration by \(g_i\), the deterministic system is described by the equations:
\begin{equation}
\begin{aligned}
\frac{\dif g_1}{\dif t} &= \frac{a_1}{1 + (g_2/K_2)^n} - \frac{g_1}{\tau} \\ \\
\frac{\dif g_2}{\dif t} &= \frac{a_2}{1 + (g_1/K_1)^m} - \frac{g_2}{\tau}
\end{aligned}
\end{equation}
where \(a_1 = 156\), \(a_2 = 30\), \(n = 3\), \(m = 1\), \(K_1 = 1\), \(K_2 = 1\), \(\tau = 1\).
In our stochastic model, our chemical reactions take the form of a birth-death process, the propensity functions of which are taken from the above differential equations:
\begin{equation}
\begin{aligned}
\varnothing \xrightarrow{k_1(g_2)} g_1 \xrightarrow{k_0} \varnothing \\
\varnothing \xrightarrow{k_2(g_1)} g_2 \xrightarrow{k_0} \varnothing 
\end{aligned}
\end{equation}
where \( k_0 = 1/\tau \), \( k_1(g_2) = a_1/[1+(g_2/K_2)^n] \), \( k_2(g_1) = a_2/[1+(g_1/K_1)^m] \).

\subsubsection{WE Parameters}

For this system we implemented 2-dimensional bins: 15 along \(g_1\) and 31 along \(g_2\), for a total of 465 bins.  The bins along the \(g_1\) coordinate were of unit width on the interval \([0,10]\), and then of width 10 on the interval \([10,50]\), with one additional bin on \([50,\infty]\).  The bins along the \(g_2\) coordinate were of unit width on the interval [0,30], with one additional bin on \([30,\infty]\).  

The WE data in Fig.~\ref{fig:e-switchMFPT} was generated using 16 trajectory segments per bin, a time-step \(\tau = 0.1 \: \si{\second}\), and run for 500 iterations of that time-step, with reweighting events applied every 100 iterations.  Fig.~\ref{fig:e-switchMFPT} shows six independent simulations using these parameters, as well as MLE statistics from our brute-force computations.  Were all the bins full at all iterations, WE would compute, for each of the six runs, 3,720,000 trajectory segments of length 0.1 seconds each, which is equivalent in cost to running 372,000 seconds of brute-force dynamics.  In our case, most of the bins never get populated; we computed dynamics for 148,855, 149,516, 148,940, 147,351, 146,804, and 149,765 segments in the six different runs.  In toto, this is equivalent to 89,123.1 seconds of brute-force dynamics.

\subsubsection{Results}

Even the state-space of this two-species stochastic system is too large to solve exactly, necessitating the use of brute-force simulation as a baseline comparison.  A brute-force computation was performed using the SSA as implemented in BNG. 753 simulations of \(10^6\) seconds each were run, and using an exponential distribution of MFPTs, the MLE (see Eq.\ \ref{eq:mld}) of the mean and standard error of the mean, \(\mu_\text{MLE}\) and \(\sigma_\mu\), were found to be \(1.3 \times 10^6\) seconds and \(6.5 \times 10^4\) seconds respectively.

\begin{figure}[htbp]
   \centering
   \includegraphics[width=8.5cm]{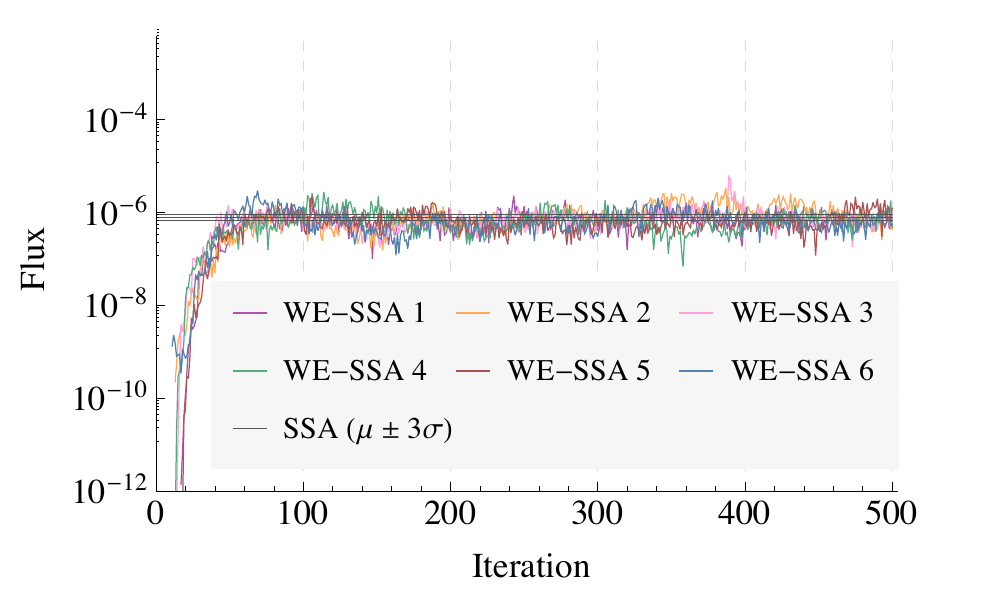} 
   \caption[Epigenetic Switch Flux]{Measurements of probability flux into the target state for the epigenetic switch system.  Six independent WE simulations are plotted, as well as the 3-\(\sigma\) confidence interval for the brute-force data, which is from 753 trajectories of \(10^6\) seconds each.  The inverse of the flux gives the mean first passage time by Eq.\ \protect{\ref{eqn:mfpt}}.}
   \label{fig:e-switchMFPT}
\end{figure}

The WE results are plotted against the brute-force values in Fig.~\ref{fig:e-switchMFPT}, where we have used the relation \(\text{MFPT = 1/\text{flux}}\) (Eq.~\ref{eqn:mfpt}) to plot the steady-state flux that brute-force predicts.  We plot the net flux entering the target state as the simulation progresses, because this is what WE measures directly; we can infer the MFPT using the above relation.  Taking the mean of each of the six WE runs after the simulation is in steady-state (we discard the first 100 iterations), and treating each of these means as an independent data point, WE gives a combined estimate for the MFPT of \(1.3 \times 10^6 \pm 3 \times 10^4\) seconds at 1-\(\sigma\).

WE is able to find an estimate of the MFPT with greater precision than brute-force, using the equivalent of 89,123.1 seconds of brute-force dynamics.  The brute-force estimate uses \(753 \times 10^6\) seconds of dynamics, yielding a speedup by a factor of \(E \sim 10^4\) when using WE compared to brute-force.

WE is also able to quickly attain an efficient rough estimate of the MFPT.  A brute-force estimate of the MFPT would require, optimistically, computing an amount of dynamics on the order of the MFPT itself (\(\sim\)\(10^6 \: \si{\second}\)).  In the six different simulations, WE's estimate of the MFPT is within \(50\%\) of the brute-force value after \(\{52,44,37,40,43,42\}\) iterations of WE simulation, at which point \(\{10238,8400,6177,6819,7141,7750\}\) trajectory segments have been propagated, which is equivalent to propagating \(\{1023.8,840.0,617.7,681.9,714.1,775.0\}\) seconds of brute-force dynamics, the mean of which is approximately 775.  Thus we find a mean \(E_{50\%} \approx 1.3 \times 10^6 / 775 \approx 1725\).

\subsection{Fc{\textepsilon}RI-Mediated Signaling}

\subsubsection{Model}

To demonstrate the flexibility of the WE approach, we applied it to a signaling model that is, to our knowledge, considerably more complex than any other biochemical system to which rare event sampling techniques have been applied. The reaction network in this model (see supplementary material [URL will be inserted by AIP] fceri\_ji.bngl) contains 354 chemical species and 3680 chemical reactions~\cite{Faeder2003}.

This model describes association, dissociation, and phosphorylation reactions among four components: the receptor Fc{\textepsilon}RI, a bivalent ligand that aggregates receptors into dimers, and the protein tyrosine kinases Lyn and Syk. The model also includes dephosphorylation reactions mediated by a pool of protein tyrosine phosphatases.   These reactions generate a network of 354 distinct molecular species. The model predicts levels of association and phosphorylation of molecular complexes as they vary with time, ligand concentration, concentrations of signaling components, and genetic modifications of the interacting proteins.

\subsubsection{WE Parameters}

The WE data in Fig.~\ref{fig:reksykpsPDF} was generated using 60 bins of unit width, 100 trajectory segments per bin, a time-step \(\tau = 0.6\: \si{\second}\), and run for 100 iterations of that time-step, with no reweighting events.  The brute-force data is from 1484 brute-force runs of 60 seconds each, which is equivalent to the dynamics time employed in attaining the WE data.  No attempt was made to optimize sampling times or bin widths in WE.

\subsubsection{Results}

\begin{figure}[htbp]
   \centering
   \includegraphics[width=8.5cm]{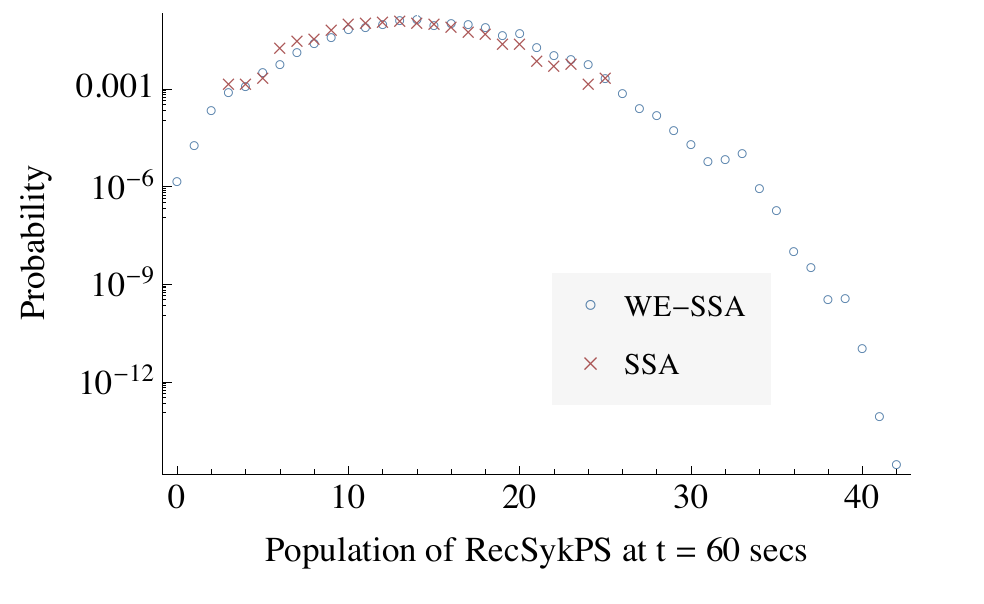} 
   \caption[Fc{\textepsilon}RI-Mediated Signaling PDF]{Comparison of WE and SSA for the Fc{\textepsilon}RI signaling model, which has 354 reactions and 3680 chemical species. The probability distribution is shown for the system reaching a specified level of Syk activation (the output of the model, which is a sum of 164 species concentrations) within one minute of system time after stimulation. Results of 1484 SSA simulations of one minute duration are compared with WE results generated with an equivalent computational effort (several CPU hours in each case).}
   \label{fig:reksykpsPDF}
\end{figure}

Fig.~\ref{fig:reksykpsPDF} shows the probability distribution of activated receptors in the Fc{\textepsilon}RI-Mediated Signaling model at time \(t = 60 \: \si{\second}\).  The brute-force SSA approach is unable to sample out to likelihoods much below \(\sim\)\(10^{-3}\), while WE gets clean statistics for likelihood values down to \(\sim\)\(10^{-15}\), for an estimated improvement in efficiency \(E \sim 10^{12}\).

\section{Discussion}

We applied the weighted ensemble (WE)~\cite{Huber1996,Zhang2007,Zhang2010,Bhatt2010,Zwier2011,Adelman2013,Lettieri2012,Zwier2013} approach to systems-biology models of stochastic chemical kinetics equations, implemented in BioNetGen~\cite{Blinov2004,Faeder2009}.  Increases in computational efficiency on the order of \(10^{20}\) were attained for a simple system of biological relevance (the enzymatic futile cycle), and on the order of \(10^{12}\) for a large systems-biology model (Fc{\textepsilon}RI), with 354 species and 3680 reactions.

WE is easy to understand and implement, statistically exact~\cite{Zhang2010}, and easy to parallelize.  It can yield long-timescale information such as mean first passage times (MFPTs) from simulations of much shorter length.  As in prior molecular simulations~\cite{Bhatt2010,Zhang2007}, WE has been demonstrated to increase computational efficiency by orders of magnitude for models of non-trivial complexity, and offers perfect (linear) parallel scaling.  It appears that WE holds significant promise as a tool for the investigation of complex stochastic systems.

Nevertheless, a number of additional points, including limitations of WE and related procedures, merit further discussion.

\subsection{Strengths of WE}

Beyond the efficiency observed for the systems studied here, the WE approach has other significant strengths.  Weighted ensemble is easy to implement: it examines trajectories at fixed time-intervals, and its implementation as scripting-level code makes it amenable to using any stochastic dynamics engine to propagate trajectories.  WE also parallelizes well, and can take advantage of multiple cores on a single machine, or across many machines on a cluster; Zwier et al.\ have successfully performed a WE computation on more than 1,000 cores on the Ranger supercomputer~\cite{Zwier2013}.  Additionally, WE trajectories are unbiased and follow the natural dynamics of the system. WE also yields full probability distributions, and can find mean first passage times (MFPTs) and equilibrium properties of systems.

\subsection{Comparison to Other Approaches}

WE is most similar in spirit to recent versions of forward flux sampling (FFS)~\cite{Allen2009} and non-equilibrium umbrella sampling (NEUS)~\cite{Dickson2009}.  All  of these methods divide up state-space into different regions, and are able to merge and split trajectories so as to enhance the sampling of rare regions of state-space.  The approaches described differ slightly in the way the splitting and merging of trajectories is performed.  WE also differs from FFS in that WE does not have to catch trajectories in the act of crossing a bin boundary; instead WE checks, at a prescribed time step, in which bin a trajectory resides.  This can be advantageous in that no low-level interaction with the dynamics engine/software is required in WE.

The central hurdle to improving efficiency using accelerating sampling techniques such as WE, FFS, and NEUS, is to adequately divide that state-space by selecting reaction coordinates that are both important to the dynamics of interest, and that are slowly sampled by brute-force approaches.  Optimally and automatically dividing and binning the state-space is, to our knowledge, an open problem, and one that, for complex systems, where a target state is unknown, is not always a straightforward one to solve, though adaptive strategies have been suggested~\cite{Zhang2010,Adelman2013,Bhatt2012}.

The wSSA approach~\cite{Kuwahara2008} differs from the above approaches.  It does not use a state-space approach, but rather uses importance sampling to bias and then unbias the dynamics.  WE seems to have comparable performance to wSSA for systems to which both can be applied.  Since wSSA biases/unbiases reaction rates, while WE divides state-space, the advantage of one over the other may be situation-dependent.  The ease of implementation of the WE framework would appear to scale better with model complexity than current versions of wSSA, though for very small models wSSA may outperform WE in measuring select observables.

A limitation common to accelerated sampling techniques used to estimate non-equilibrium observables is the system-intrinsic timescale: ``\(t_b\)'', or the ``event duration'' time~\cite{Zhang2007a}.  This timescale represents the time it takes for realistic trajectories to ``walk'' from one state to another, excluding the waiting time prior to the event.  The event duration is often only a fraction of the MFPT, since it is the \emph{likelihood} of walking this path that is low; the time to actually walk the path is often quite moderate.  That is, the waiting time in an initially metastable state can greatly exceed \(t_b\).  WE excels at overcoming the low likelihood of a transition, but no accelerated sampling technique can overcome \(t_b\).

Finally, it should be noted that all state-space methods that branch trajectories, including WE, produce correlated trajectories, due to the splitting/merging events.  While such correlations do not appear to have impeded the application of WE to the systems investigated here, future work will aim to quantify their effects and reduce their potential impact.  The present work accounted for correlations by analyzing multiple fully independent WE runs.

\subsection{Future Applications}

Beyond potential applications to more complex stochastic chemical kinetics models, the weighted ensemble formalism could be applied to spatially heterogeneous systems.  WE should be able to accelerate the sampling of models such as those generated by MCell \cite{Stiles2001,Kerr2008,Stiles1996} or Smoldyn \cite{Andrews2012}, perhaps using three-dimensional spatial bins.

It may be possible to integrate WE with other methods.  We note that the state-space dividing approaches of a number of methods (forward flux~\cite{Allen2005,Allen2006a,Allen2006,Allen2009}, non-equilibrium umbrella sampling~\cite{Dickson2009,Warmflash2007}, and weighted ensemble~\cite{Huber1996,Zhang2007,Zhang2010,Bhatt2010,Zwier2011,Adelman2013,Lettieri2012,Zwier2013}), since they are dynamics-agnostic, could be combined with other methods that accelerate the dynamics engine itself, such as the \(\tau\)-leaping modification of Gillespie's SSA and its many variants and improvements~\cite{Harris2006,Tian2004,Cao2006,Gillespie2001}, to yield multiplicative increases in runtime speedup.

More speculatively, WE could be combined with parallel tempering methods~\cite{Hansmann1997,Sugita1999,Earl2005}.  WE accelerates the exploration of the free-energy landscape at a given temperature, and since it does not bias dynamics, the trajectories it propagates could be suitable for replica exchange schemes.

For complex models where exploring the state-space via brute-force is prohibitively expensive, WE could also be employed to search for bistability, or in a model-checking capacity \cite{Clarke1999,Heiner2008,Degano2009} to search for pathological states.

\section{Acknowledgments}

We gratefully acknowledge funding from NSF grant MCB-1119091, NIH grant P41 GM103712, NIH grant T32 EB009403, and NSF Expeditions in Computing Grant (award 0926181). We thank Steve Lettieri, Ernesto Suarez, and Justin Hogg for helpful discussions.


%

\end{document}